\documentclass[aps,prl,nofootinbib,twocolumn,superscriptaddress,preprintnumbers,nofootinbib]{revtex4-1}
\pdfoutput=1

\usepackage{amssymb}
\usepackage{amsmath}
\usepackage{epsfig}
\usepackage{hyperref}
\usepackage{breakurl}
\usepackage{color}

\makeatletter
\def\simgt{\mathrel{\lower2.5pt\vbox{\lineskip=0pt\baselineskip=0pt
           \hbox{$>$}\hbox{$\sim$}}}}
\def\simlt{\mathrel{\lower2.5pt\vbox{\lineskip=0pt\baselineskip=0pt
           \hbox{$<$}\hbox{$\sim$}}}}
\makeatother

\newcommand{\be}{\begin{equation}}
\newcommand{\ee}{\end{equation}}
\newcommand{\bea}{\begin{eqnarray}}
\newcommand{\eea}{\end{eqnarray}}
\newcommand{\eq}[2]{\be\begin{aligned}#1 \label{#2}\end{aligned}\ee}

\newcommand{\Eq}[1]{Eq.~\eqref{#1}}

\newcommand{\LL}{\mathcal{L}}

\newcommand{\OO}{\mathcal{O}}
\newcommand{\DD}{\mathcal{D}}
\newcommand{\KK}{\mathcal{K}}

\newcommand{\nn}{\nonumber}

\def\Section#1{\vskip .05 cm \noindent {\it #1}}

\begin{document}

\title{Hidden conformal invariance of scalar effective field theories}

\author{Clifford Cheung}
\affiliation{Walter Burke Institute for Theoretical Physics,
California Institute of Technology,\\ Pasadena, CA 91125, USA}

\author{James Mangan}
\affiliation{Walter Burke Institute for Theoretical Physics,
California Institute of Technology,\\ Pasadena, CA 91125, USA}

\author{Chia-Hsien Shen}
\affiliation{Mani L. Bhaumik Institute for Theoretical Physics,
Department of Physics and Astronomy,\\ UCLA, Los Angeles, CA 90095, USA}

\begin{abstract}

We argue that conformal invariance is a common thread linking several scalar effective field theories that appear in the double copy and scattering equations. For a derivatively coupled scalar with a quartic ${\cal O}(p^4)$ vertex, classical conformal invariance dictates an infinite tower of additional interactions that coincide exactly with Dirac-Born-Infeld theory analytically continued  to spacetime dimension $D=0$.  For the case of a quartic ${\cal O}(p^6)$ vertex, classical conformal invariance constrains the theory to be the special Galileon in $D=-2$ dimensions.  We also verify the conformal invariance of these theories by showing that their amplitudes are uniquely fixed by the conformal Ward identities. In these theories, conformal invariance is a much more stringent constraint than scale invariance.

\end{abstract}

\maketitle

\Section{Introduction.}
The modern scattering amplitudes program has exposed an array of extraordinary theoretical structures which include the double copy~\cite{KLT, BCJ1, BCJ2, color_kinematics_review}, scattering equations \cite{CHY1, CHY2,CHY3,CHY4}, and novel reformulations of amplitudes as polyhedra \cite{Arkani-Hamed:2013jha, Arkani-Hamed:2017mur}.
Developing these theoretical structures has also led to important applications.
For instance, via the double copy procedure, gravity's highly complex amplitudes can be obtained by ``squaring'' much simpler amplitudes from gauge theory. This simplification sits at the heart of the recent state-of-art calculation of the black hole binary Hamiltonian at third post-Minkowskian order~\cite{Bern:2019nnu,Bern:2019crd}.
Therefore, it cannot be overemphasized how important it is to understand the origins of these novel structures and to carve out the space of theories that enjoys these properties.

Curiously, the same set of theories emerges again and again when studying the double-copy and scattering equations.
This set includes well-known theories like gravity and Yang-Mills (YM) in addition to a variety of scalar theories such as the biadjoint scalar (BS), the nonlinear sigma model (NLSM), Dirac-Born-Infeld (DBI) theory\footnote{In this paper, we consider DBI theory in flat space, rather than the conformal DBI, which describes a brane in an anti-de Sitter background.}, and the special Galileon~\cite{Galileon, EFTFromSoft, CHY4}.
These scalar theories can be viewed as the cousins of YM and gravity and sometimes serve as simple toy models to decode mysterious properties like the double copy~\cite{Cheung:2016prv}.
Gravity, YM, and these scalar theories are also exceptional in that their interactions are fully fixed by economical principles such as  Lorentz invariance~\cite{Benincasa:2007xk, Elvang:2013cua, Cheung:2017pzi}, gauge invariance~\cite{Arkani-Hamed:2016rak}, soft theorems~\cite{EFTFromSoft,EFTTable,Rodina:2016jyz,Rodina:2018pcb,Elvang:2018dco,Padilla:2016mno,Low:2019ynd,Low:2017mlh,Low:2018acv,Cheung:2018oki,Kampf:2019mcd},  color-kinematics duality~\cite{BCJ1, BCJ2,KLT,color_kinematics_review,CHY4,Chen:2013fya,Carrasco:2016ldy,XYZ,Cheung:2017yef}, unifying relations~\cite{Cheung:2017ems}, ultraviolet behavior \cite{Rodina:2016mbk,Carrasco:2019qwr,Rodina:2020jlw}, or symmetry \cite{Hinterbichler_2015,Novotny:2016jkh,Bogers:2018zeg,Bogers:2018zeg,Roest:2019oiw,Roest:2019dxy,Roest:2020vny}, depending on the theory in question.
Although the details of these constructions will not be important to this paper, they motivates us to ask what physical property unites this disparate theories?

We propose that there is an underlying \emph{symmetry} connecting these theories:  conformal invariance.
For the appropriate critical spacetime dimension $D$, the coupling constant is dimensionless and classical scale invariance is trivially ensured for
BS theory ($D=6$), YM theory ($D=4$), gravity ($D=2$) and the NLSM ($D=2$).  
Notably, YM and the NLSM are curiously similar in their respective critical dimensions, e.g., both exhibit asymptotic freedom and a gapped spectrum. Rather enticingly, versions of these theories which are conformally invariant at the quantum level also expose integrable properties.

While these facts may be incidental, they beg the question of whether DBI and the special Galileon have special conformal properties.  Indeed, we will show that these scalar effective field theories (EFTs) are the unique derivatively coupled, classical conformally invariant theories in $D=0$ and $D=-2$, respectively.  While these are clearly unphysical choices for the spacetime dimension, our analysis is well-defined provided we work in general $D$ throughout and only analytically continue to these particular values at the very end.\footnote{Note a very interesting recent conjecture of conformal invariance of graviton and YM amplitudes in arbitrary dimension $D$~\cite{Loebbert:2018xce}, later proven in Ref.~\cite{Nutzi:2019ufl}.}

A corollary of our result is that the tree-level scattering amplitudes in these EFTs are annihilated by the generators of the conformal group.  We then show how the conformal Ward identities---together with Lorentz invariance, locality, factorization, and the leading Adler zero \cite{Adler}---are sufficient to uniquely bootstrap these amplitudes, confirming via an amplitudes analysis that the corresponding EFTs are fixed by classical conformal invariance.

In addition, our results show that scale invariance does not imply conformal invariance in the peculiar $D=0$ and $D=-2$ cases we will discuss.
Typically, scale invariance implies conformal invariance in numerous contexts \cite{1970AnPhy..59...42C,Zamolodchikov:1986gt,Polchinski:1987dy,Komargodski:2011vj,Luty:2012ww,Dymarsky:2013pqa,Dymarsky:2014zja,Dymarsky:2015jia} when principles like unitarity are assumed. It is unclear whether these assumptions hold in the unphysical dimension $D$ here. In fact, our results are concrete examples where conformal invariance imposes further constraints beyond scale invariance.

\medskip

\Section{Lagrangians from Conformal Invariance.}
\label{LagrangianSection}
An obvious necessary condition for conformal invariance is scale invariance.  Scale invariance requires that all coupling constants of the theory are dimensionless in a given critical dimension $D$.  \footnote{Free theories such as Maxwell (see Ref.~\cite{NakayamaMaxwell}) and Klein-Gordon can be scale invariant outside of their naive critical dimensions.}
Following Ref.~\cite{EFTFromSoft}, we define a power counting parameter $\rho$ which characterizes the number of derivatives per interaction for a derivatively coupled scalar field $\phi$.  A generic vertex takes the form \footnote{We will assume manifest locality so that no derivatives appear with negative powers in $\LL$.}
\begin{align}
\label{RhoPowerCounting}
 (\partial \phi)^2 (g \partial^\rho \phi)^{n-2},
\end{align}
where $g$ is the coupling constant and the precise placement of derivatives, i.e., which derivative acts upon which field, is schematic and should be disregarded.
Symmetries generally relate interaction vertices of the same $\rho$, since by dimensional analysis these terms can destructively interfere in scattering amplitudes.  Scale invariance implies that $g$ is dimensionless.  So, in the critical dimension, $D$ and $\rho$ are related to each other by
\begin{equation}
	-\rho = \Delta = \frac{D - 2}{2},
	\label{eq:power_counting}
\end{equation} 
where we have used that the field $\phi$ has dimension $\Delta = (D-2)/2$.
An important feature is that in the critical dimension $D\le 2$, we have $\rho\ge 0$ and therefore scale invariance alone still permits an infinite tower of marginal interactions.  However, as we will see shortly, the additional assumption of conformal invariance will actually fix this tower uniquely for derivatively coupled scalars.  
In particular, scale invariance merely implies that $T\equiv T^\mu_{\;\mu} =dJ$ for some virial current $J$, while conformal invariance imposes the additional constraint that the virial current is conserved, so $T =dJ =0$. 

As is well-known, however, the energy-momentum tensor is only defined modulo improvement terms which are identically conserved,
so conformal invariance requires that $T=0$ up to this ambiguity.
A mechanical algorithm to enumerate these improvement terms is to couple the theory to a background metric, 
\eq{
	\hat\LL = \sqrt{-g} \left( \LL + \Delta \LL  \right),
}{eq:grav_L}
including all possible minimal and nonminimal gravitational couplings.
Since the energy-momentum tensor is the first variation of the background metric, we need to only include nonminimal gravitational interactions which are linear in the Riemann tensor.  Higher powers will only contribute to the second variation and higher.  Since the linear variation of Riemann has two derivatives in it, the resulting energy-momentum tensor has a trace $T$ which is corrected by some improvement operator of the form  $\partial \partial L$ for some local rank two tensor $L$.  Hence, the most general statement of conformal invariance is that $T=\partial  \partial L$.\footnote{See Ref.~\cite{Nakayama:2013is} for a pedagogical review and references therein.}


For our analysis, we begin by constructing a general ansatz Lagrangian for a derivatively coupled scalar field $\phi$ with interactions at a fixed value of $\rho$.   Much like in dimensional regularization, we work in general dimensions such that the variable $D$ only appears at the very end through $\eta^\mu_{\;\; \mu}=D$.  We thus ignore all Gram determinant or evanescent effects since these are of course ill-defined for unphysical dimension $D$ anyway. We then constrain the coefficients of the ansatz Lagrangian using conformal invariance.  

\medskip
\Section{Nonlinear Sigma Model.}
\label{sec:Lagrangian_pion}
As a warm up, consider the case of $\rho=0$, which describes a theory of scalars with at most two derivatives per interaction. 
This analysis is simple but will serve as a template for more complicated EFTs.  
The most general two-derivative Lagrangian is\footnote{We work in mostly plus signature throughout.}
\eq{
\LL = -\frac{1}{2} \partial^\mu \phi^i  \partial_\mu \phi^{j} K_{ij}, 
}{}
where $i,j$ are internal (target space) indices and $K_{ij}(\phi)$ is field dependent. 
We will compute the energy-momentum tensor from the coupling to a metric.
We couple this theory to a background metric via
\eq{
\hat\LL = \sqrt{-g} \left( \LL + R \, W  \right),
}{eq:NLSM_L}
where $\LL$ above is properly covariantized and the arbitrary function $W(\phi)$ parameterizes the improvement terms induced by nonminimal coupling to the Ricci scalar.
The energy-momentum tensor is obtained from the first variation of the metric about flat space, $T^{\mu\nu} = 2\frac{\delta S}{\delta g_{\mu\nu}}$, so
\begin{align}
\label{eq:TNLSM}
T =&  -\frac{1}{2}\partial^\mu \phi^i \partial_\mu \phi^j K_{ij}(D-2) \nonumber\\
&- 2(D-1)(\partial^\mu \partial_\mu \phi^i W_i + \partial^\mu \phi^i \partial_\mu \phi^j W_{ij}),
\end{align}
where $W_i = \tfrac{d W}{d\phi_i}$ and $W_{ij}= \tfrac{d^2 W}{d\phi_i d\phi_j}$. 
Thus, in the absence of improvement terms, any two-derivative theory is classically conformal in $D=2$.   In this case, conformal invariance places no restriction on $K_{ij}$ and is identical to scale invariance.

Another well-known example is free theory, where $K_{ij} = \delta_{ij}$.
Inserting the equations of motion $\Box \phi^i=0$ into Eq.~\eqref{eq:TNLSM}, we obtain
\begin{align}
\label{eq:T_free}
T =&  -\frac{1}{2} \partial^\mu \phi^i \partial_\mu \phi^j \left[
 (D-2) \delta_{ij}
+4(D-1) W_{ij}
\right],
\end{align}
so for $W_{ij} = - \tfrac{D-2}{4(D-1)} \delta_{ij}$ we obtain a set of conformally-coupled scalars in any dimension.  Note that the first term on the right-hand side of \Eq{eq:T_free} is equal to $ -\frac{1}{4} \partial^\mu\partial_\mu\left[
(\phi^{i})^2 (D-2)  \right]$ on the support of the free equations of motion.  Consequently, in the absence of improvement terms, the trace of the energy-momentum tensor is of the form $T = \partial_\rho \partial_\sigma L^{\rho\sigma}$, as expected for a conformally invariant theory.

\medskip
\Section{Dirac-Born-Infeld Theory.}
\label{LagrangianDBISection}
We now turn to the case of $\rho=1$, which is scale-invariant in $D=0$.
For a derivatively coupled scalar, the Lagrangian is an arbitrary polynomial in $X=(\partial \phi)^2$.\footnote{Working with functions of $X$ is the simplest way to satisfy Lorentz invariance and power counting.  However, introducing a scalar multiplet with a more complex derivative structure could lead to more elaborate brane theories such as the multi-field DBI appearing in \cite{EFTTable}.}
Coupling this theory to a background metric, we obtain
\eq{
\hat{\LL} = \sqrt{-g} \left(\LL + R\, A \, \phi^2 + R^{\mu\nu} B\, \phi^2 \nabla_\mu\phi \nabla_\nu\phi  \right),
}{eq:DBIHat}
where $A(X)$ and $B(X)$ are undetermined functions of $X$.
{\it A priori}, one can add nonminimal couplings to the Riemann tensor but these all vanish by antisymmetry given the number of derivatives.
The trace of the energy-momentum tensor is
\begin{align}
\label{eq:T_DBI}
T =& - 2 \LL' X + D \LL + (2-D) \partial^\mu\partial^\nu \left( \phi^2 \partial_\mu \phi \partial_\nu\phi B\right) \nonumber\\
&+2\partial^\mu \partial_\mu \left[ \phi^2 \left\{ (1-D)A - \frac{1}{2} B X\right\}\right],
\end{align}
where the prime denotes differentiation with respect to $X$.

For classical conformal invariance, $T=0$ modulo the equations of motion,
\begin{align}
\Box \phi = -2\frac{\LL''}{\LL'} Y_\mu Y_\nu Z^{\mu\nu} ,
\end{align}
where $Y_\mu = \partial_\mu \phi$ and $Z_{\mu\nu}=\partial_\mu\partial_\nu\phi$.
Plugging this into \Eq{eq:T_DBI}, we find that 
\begin{align}
T&= \sum_{i=1}^6 c_i (X){\cal O}_i
\end{align}
can be expanded in a basis of six tensor structures,
\begin{align}
{\cal O}_i = \{&
1,~
\phi Y_\mu Y_\nu Z^{\mu\nu} ,~
\phi^2 (Z^{\mu\nu} )^2,~
\phi^2 Y_\mu Y_\nu Y_\rho W^{\mu\nu\rho},\nonumber\\
&
\phi^2 (Y_\mu Z^{\mu\nu} )^2,~
\phi^2(Y_\mu Z^{\mu\nu} Y_\nu)^2  \},
\end{align}
where $W_{\mu\nu\rho}=\partial_\mu\partial_\nu\partial_\rho \phi$ and the coefficients $c_i(X)$ are
\begin{align}
\label{eq:diff_eq_DBI_1}
c_1&= 2X(2A + B X- \LL') \\
\label{eq:diff_eq_DBI_2}
c_2&=
4(4A'+B+2B'X) -4(2A+3BX)\frac{\LL''}{\LL'}\\
c_3&=2(2 A'- B' X)\\
c_4&=4B' -4(2A'+B-B' X)\frac{\LL''}{\LL'}\\
c_5&= 4(2A''+B'-B''X)-8(2A'+B-B'X)\frac{\LL''}{\LL'}\\
c_6&= 8 B'' -16 B' \frac{\LL''}{\LL'} -8 (2A'+B-B'X)\frac{\LL'''}{\LL'}\nonumber\\
& \phantom{{}=}+8 (2A'+2B-B'X)\left(\frac{\LL''}{\LL'}\right)^2 \,.
\end{align}
Treating each $\OO_i$ as independent, we find that $c_i=0$, yielding a system of differential equations for $\LL$, $A$, and $B$.
First, we solve $c_1=0$ for $A$.  Plugging $A$ and $A'$ into $c_2=0$ gives an {\it algebraic} expression for $B$ in terms of derivatives of $\LL$.  Finally, inserting $A$ and $B$ and their derivatives into $c_3=0$ yield
\eq{
 \LL' \LL''' &= 3 \LL''^2,
}{}
from which we obtain the general solution,
\eq{
\LL &= -\frac{1}{g}\sqrt{1+gX}+\lambda\\
A &= -\frac{gX+2}{8\sqrt{1+gX}}\\
B &= \frac{g}{4\sqrt{1+gX}} ,
}{eq:L_DBI}
which also solves the remaining equations.
Here the decay constant $g$ and cosmological constant $\lambda$ arise as constants of integration.
Remarkably, we narrow down to this particular solution from a class of scale-invariant theories, showing the former is much stronger than the latter in $D=0$.
We thus arrive at a main result of this paper: DBI is the unique conformally invariant, derivatively coupled scalar in $D=0$.

\medskip
\Section{Special Galileon.}
Next, let us move on to theories with $\rho=2$, which are scale invariant in $D=-2$. 
We choose a basis for a derivatively coupled scalar where the $n$-point interaction vertex takes the form $c^{\mu_1 \dots \mu_{2n-2}}_n Y_{\mu_1} Y_{\mu_2} Z_{\mu_{3} \mu_{4}}\dots Z_{\mu_{2n-3} \mu_{2n-2}}$, where $c_n$ is an arbitrary constant tensor built from the flat space metric and numerical coefficients.   As before, we promote this theory to couple with a background metric and then include all possible improvement terms built from Riemann contracted with derivatives of the scalars, taking the schematic forms $R \phi^2 Z^{n-2}$, $R \phi Y^2 Z^{n-3}$, and $R Y^4 Z^{n-4}$.

Setting $T=0$ on the support of the equations of motion in $D=-2$, we derive constraints on the interaction coefficients though six point.  Conformal invariance fixes many but not all of the couplings in the ansatz Lagrangian.  Nevertheless, by computing the scattering amplitudes in the resulting theory via Feynman diagrams, we discover that they coincide {\it exactly} with those of the special Galileon.  Hence, the unfixed Lagrangian parameters all evaporate on-shell and can be eliminated by an appropriate field redefinition.

In fact, through a suitable choice of the unfixed parameters, the Lagrangian can be brought to the original representation of the special Galileon \cite{Hinterbichler_2015},
\begin{align}
\LL =& -\frac{1}{2}X \Big\{  1 - \frac{1}{3!} \left( [Z]^2-[Z^2] \right)+\frac{1}{5!}\big([Z]^4-6[Z]^2[Z^2] \nn \\
&+3[Z^2]^2+8[Z][Z^3]-6[Z^4]\big)\Big\} +  \dots,
\label{eq:L_sGal}
\end{align}
where the square brackets denote a trace over spacetime indices $[Z^n] = Z^{\mu_1}_{\;\;\;\;\mu_2}Z^{\mu_2}_{\;\;\;\;\mu_3}\dots Z^{\mu_{n}}_{\;\;\;\;\mu_1}$.    The freedom of unfixed couplings can also be used to put the improvement terms in a form that 
depends only on the Ricci tensor,
\begin{align}
\Delta \LL&= \phi^2 \Big(-\frac{1}{6}[R] -\frac{1}{72} [R] [Z^2] + \frac{1}{12} [RZ^2]-\frac{1}{20} [RZ^4] \nonumber\\
& \phantom{{}=}+\frac{1}{40} [RZ^2][Z^2]-\frac{1}{90} [RZ][Z^3]+\dots \Big),
\end{align}
which closely mimics those of DBI in \Eq{eq:DBIHat}.
While it is computationally difficult to extend these results to higher point, this pattern will almost certainly continue.  We leave the question of conformal invariance to all orders for future work.

\medskip
\Section{Scattering Amplitudes from Conformal Invariance.}
Conformal invariance can be enforced at the level of scattering amplitudes rather than the Lagrangian.  This has the distinct advantage of trivializing equations of motion and eliminating ambiguities arising from field redefinitions.  Here we consider two types of amplitudes constraints which both imply and are implied by conformal invariance.  

The first constraint requires coupling the scalar EFT in question to an additional dilaton degree of freedom, $\tau$.   Since the dilaton couples via $\tau T$ and conformal invariance implies that $T = \partial\partial L$, the single-dilaton amplitude exhibits a \emph{double Adler zero} in the soft limit, 
\eq{
A_{n+1}(q,p_1,\dots,p_n)|_{q\rightarrow 0} \sim \mathcal{O}(q^2),
}{eq:dilaton_double}
where $q$ is the dilaton momentum.  To reach this conclusion, one must in general be careful about soft propagator poles spoiling the double Adler zero.  However, this is not a problem in a theory of derivatively coupled scalars since the on-shell three-point amplitude vanishes identically due to kinematics.

Notably, the converse proposition is also true: the double Adler zero in \Eq{eq:dilaton_double} implies conformal invariance.  To understand this, consider $A_{n_{\rm min}+1}$ for the smallest possible number of EFT scalars $n_{\rm min}$ for which the amplitude is nontrivial.  By definition, $A_{n_{\rm min}+1}$ is a local interaction vertex evaluated on-shell with no internal propagators.  
The $\mathcal{O}(q^2)$ soft behavior of the dilaton implies that the lowest order interaction vertex of the dilaton in the off-shell Lagrangian is an operator of the form $\partial \partial \tau  L$, where $L$ is a local operator that depends on the EFT scalars.  Of course, this operator is ambiguous up to terms which vanish on-shell.  Crucially, however, these terms all involve either the on-shell condition for the dilaton, $\Box \tau$ or the on-shell condition for the scalar, $\Box \phi$.  The former produces contributions {\it still} of the form $\partial \partial \tau  L$, while the latter can be eliminated via a field redefinition in favor of higher order terms.  

Next, we consider $A_{n+1}$ for $n> n_{\rm min}$.  This amplitude has propagator poles, but all the singularities must factorize into lower-point dilaton amplitudes times scalar amplitudes.  On these factorization channels, there is always a double Adler zero because the lowest order dilaton interaction vertex is of the form $T=\partial \partial L$ and as discussed before, there are no on-shell three-point amplitudes. Consequently, the residual contact term in the amplitude must independently scale as $\mathcal{O}(q^2)$ and should then be added to the definition of $L$.  This argument is then repeated for higher and higher order amplitudes until we obtain $T = \partial \partial L$ to all orders. 

The above argument establishes that a double Adler zero for the dilaton implies conformal invariance.  However, the dilaton soft theorem is also equivalent to a second type of amplitudes constraint, which is the conformal Ward identity on pure scalar EFT amplitudes.
This connection has been shown in the context of gluon and graviton amplitudes~\cite{Loebbert:2018xce}.
As discussed in Ref.~\cite{DiVecchia:2015jaq}, the dilaton soft limit 
is defined by
\begin{align}
&A_{n+1}(q,p_1,\dots,p_n)|_{q\rightarrow 0} \nn \\
=& (\mathcal{D} + q^\lambda \KK_\lambda)A_{n}(p_1,\dots,{p}_n)+\mathcal{O}(q^2),
\end{align}
where we crucially set ${p}_n = -\sum^{n-1}_{j=1} p_j$ in order to ensure that the scale and conformal operators commute with momentum conservation \cite{DiVecchia:2015jaq}.
Here $\DD$ and $\KK_\lambda$ are the scale and conformal boost generators in momentum space,
\begin{align}
\DD =& -D+n\Delta+\sum_{i=1}^{n} p_{i\nu}\cdot \partial_{i,\nu} \\
\KK_\lambda =& \sum_{i=1}^{n}\left[p_{i}^\nu \partial_{i,\lambda \nu}-\frac{1}{2}p_{i\lambda}\partial^2_{i} + \Delta \partial_{i,\lambda}\right],
\end{align}
where $\partial_{i,\nu}= \partial/\partial p^{\nu}_i$,
$\partial_{i,\mu\nu}= \partial^2/(\partial p^{\mu}_i\partial p^{\nu}_i)$ and $\partial^2_{i} =\eta^{\mu\nu} \partial_{i,\mu\nu}$. 
In the appropriate critical dimension $D$, all amplitudes are trivially annihilated by $\DD$, so the double Adler zero, and hence conformal invariance, hold if and only if
\eq{
\KK_\lambda A_{n}(p_1,\dots,{p}_n) =0.
}{eq:KA}
For explicit computations, it will be convenient to recast the conformal boost operator in terms of Mandelstam invariants $s_{ij}=-2p_i\cdot p_j$ by dotting $\KK_\lambda$ with the momentum $p_l^\lambda$ of the $l$th leg~\cite{Loebbert:2018xce}, so
\begin{align}
p_l\cdot \KK =& \sum \limits_{i,j\neq i,k \neq i} \left(s_{ik}s_{lj}-\frac{1}{2} s_{jk} s_{li}\right)\partial_{s_{ij}} \partial_{s_{ik}}\nonumber\\
 &+ \Delta\,\sum\limits_{i,j\neq i} s_{jl} \partial_{s_{ij}} \,,
\end{align}
where the spacetime dimension $D$ only enters through $\Delta = (D-2)/2$. 
Note that the above representation is well-defined because the conformal boost commutes with the on-shell condition and we have already fixed ${p}_n$ to enforce momentum conservation.

We are now equipped to use \Eq{eq:KA} to ``conformally bootstrap'' the scattering amplitudes of DBI and the special Galileon.   First, let us consider the simplest case of four-point scattering of EFT scalars.  The most general ansatz for this amplitude is a linear combination of terms like
$s^{a}_{12} s^{b}_{13}$ where $a+b = 1+ \rho$.
It is straightforward to see that $p_l\cdot \KK (s^{a}_{12} s^{b}_{13})=0$ implies that $\rho= -\Delta$, which is exactly the condition of scale invariance in \Eq{eq:power_counting}.
Thus, {\it any} scale invariant four-point scattering amplitude is automatically conformally invariant.  Note that this argument is general and applies to single or multiple scalars which may or may not be derivatively coupled. This result closely mirrors enhanced soft limits \cite{EFTFromSoft,EFTTable}, which are also automatic at four point.

For higher-point scattering, we construct
an ansatz for the amplitude $A_n$ consistent with locality, factorization, Bose symmetry, and a choice of $\rho$,
\begin{align}
A_n = A_{n,\text{cont}} + A_{n,\text{fact}} 
\label{eq:amp_ansatz}
\end{align}
where $A_{n,\text{fact}}$ is the factorization contribution obtained by treating all lower point amplitudes as Feynman vertices and summing all Feynman diagrams with at least one internal propagator.  For the residual contact contribution, we define a local ansatz function $A_{n,\text{cont}}$ which will be fixed by the conformal Ward identities.\footnote{A similar approach has been taken to study spontaneously broken conformal symmetry~\cite{DiVecchia:2017uqn}.}

To bootstrap DBI, we consider a general $\rho=1$ amplitudes ansatz for derivatively coupled scalars.  As discussed previously, four-point scattering is automatically conformally invariant.  There is no odd-point scattering due to Lorentz invariance so we jump to six point, where the only allowed interaction vertex for a derivatively coupled scalar is 
\begin{align}
A_{6,\text{cont}}=  d_6 s_{12}s_{34} s_{56} + \text{perms}.
\end{align}
for an arbitrary coefficient  $d_6$ and perms stands for the remaining sum over permutations.
The condition $\KK_\lambda A_6=0$ fixes $d_6$ so that $A_6$ is precisely the DBI amplitude.
The same procedure at eight point then fixes the contact term
\begin{align}
A_{8,\text{cont}}=  d_8 s_{12}s_{34} s_{56} s_{78} + \text{perms},
\end{align}
again in such a way that exactly matches DBI.

For the special Galileon, we build an amplitudes ansatz for $\rho=2$, derivatively coupled scalars.  As before, four point is automatic, so we start at five point where there is one independent contact term. Imposing \Eq{eq:KA} fixes $A_5$ to zero.  Moving on to six point, we perform the same exercise and reproduce the scattering amplitude for the special Galileon.  The eight point amplitude is also uniquely fixed to be the special Galileon if we assume each field has at most two derivatives.\footnote{As a cross check we have used the $\mathcal{O}(q^2)$ Adler zero for the dilaton to constrain the the scalar EFT amplitudes. We find again that DBI and the special Galileon are the unique conformally invariant, derivatively coupled amplitudes in $D=0$ and $D=-2$ up to and including six-point scattering.}

It is natural to ask whether there exist other conformally invariant theories in exotic dimensions besides DBI and the special Galileon.
We have verified that no such derivatively-coupled scalar theory exists in $D=-4$, at least up to sixth order in the field. This is perfectly analogous to the nonexistence of theories with enhanced Adler zeros at $\rho=3$ beyond four point~\cite{EFTTable}.  Note that if you relax the assumption of derivative coupling, then there exist additional scalar EFTs which are conformally invariant.  An example of such a theory is the six-point contact interaction $\phi^2 \partial_\mu\phi Z^{\mu\nu} \partial_\nu X $, which is conformal all by itself in $D=0$ but does not exhibit a shift symmetry.

\medskip
\Section{Conclusions.} 
Our findings leave a number of avenues for future study.  First, since DBI and the special Galileon are fixed by conformal invariance, it would be interesting to devise new on-shell recursion relations~\cite{Britto:2005fq} which exploit this fact.  A similar approach was taken in Ref.~\cite{EFTRecursion}, where enhanced soft limits were leveraged to derive new recursion relations for these very same scalar EFTs.

Second is the question of whether conformal invariance is exhibited by higher-spin theories in the double copy, e.g., the Born-Infeld (BI) photon, whose structure is constrained through soft behavior~\cite{Cheung:2018oki}, and the gauge theory constructed in Ref.~\cite{Johansson:2017srf}.
It would be interesting to see if the latter can be conformal in $D=6$.
On the other hand, we are actually somewhat pessimistic for BI, simply because a free photon is only conformally invariant in $D=4$, while scale invariance for interacting BI requires $D=0$. That said, a more thorough analysis, including other theories with an interacting photon~\cite{Azevedo:2017lkz}, is warranted.

Third, our results suggest an intimate connection between conformal invariance of a derivatively coupled scalar and the enhanced Adler zero condition~\cite{EFTFromSoft,EFTTable}.  Here the underlying symmetry algebras~\cite{Hinterbichler_2015,Novotny:2016jkh,Bogers:2018zeg,Bogers:2018zeg,Roest:2019oiw,Roest:2019dxy,Roest:2020vny} are likely to shed light, perhaps offering a connection to extended versions of these theories~\cite{Cachazo:2016njl,Elvang:2018dco,Carrillo-Gonzalez:2019aao,Low:2019wuv,Carrasco:2016ldy}.

Last, it would be interesting to see how conformal invariance of DBI and the special Galileon might be extended beyond the classical limit, for instance, by analyzing loops or, more speculatively, through non-perturbative means such as the conformal bootstrap analytically continued to exotic spacetime dimension.

\Section{Acknowledgments.}
We employ the symbolic manipulation package \texttt{xAct} \cite{xact} for numerous computations in this work.
C.C. and J.M. are supported by the DOE under grant no. DE- SC0011632 and by the Walter Burke Institute for Theoretical Physics.
C.-H.S. is supported by the Mani L. Bhaumik Institute for Theoretical Physics.

\bibliography{BootstrapFromConformal}

\end{document}